\def\eq#1{Eq.(\ref{#1})}

\documentclass[amsmath,amssymb,showkeys, showpacs, superscriptaddress]{revtex4}
\usepackage{graphicx}
\usepackage{amsthm}
\usepackage{amsmath}
\usepackage{amssymb}
\usepackage{array}
\usepackage{booktabs}
\usepackage{multirow}
\usepackage{color}
\newtheorem{theorem}{Theorem}[section]

\theoremstyle{definition}

\parskip \baselineskip

\begin{document}

\hspace*{5 in}CUQM-156\\
\vspace*{0.4 in}
\title{Exact and approximate solutions of Schr\"odinger's equation with hyperbolic double-well potentials}
\author{Richard L. Hall}
\email{richard.hall@concordia.ca}
\affiliation{Department of Mathematics and Statistics, Concordia University,
1455 de Maisonneuve Boulevard West, Montr\'eal,
Qu\'ebec, Canada H3G 1M8}

\author{Nasser Saad}
\email{nsaad@upei.ca}
\affiliation{School of Mathematical and Computational Sciences,
University of Prince Edward Island, 550 University Avenue,
Charlottetown, PEI, Canada C1A 4P3.}

\begin{abstract}
\noindent  Analytic and approximate solutions for the energy eigenvalues generated by the hyperbolic potentials $V_m(x)=-U_0\sinh^{2m}(x/d)/\cosh^{2m+2}(x/d),\,m=0,1,2,\dots$ are constructed. A byproduct of this work is the construction of polynomial solutions for the confluent Heun equation along  with necessary and sufficient conditions for the existence of such solutions based on the  evaluation of a three-term recurrence relation. Very accurate approximate solutions for the general problem with arbitrary potential parameters are found by use of the {\it asymptotic iteration method}.

\end{abstract}

\maketitle
\section{Introduction}\label{intro}
\noindent We study the one-dimensional Schr\"odinger equation
\begin{align}\label{eq1}
-\dfrac{\hbar}{2\mu}\dfrac{d^2\psi}{dx^2}+V(x)\psi=E\psi,\qquad V(x)=-U_0\dfrac{\sinh^4(x/d)}{\cosh^6(x/d)},\qquad \psi(\pm \infty)=0,
\end{align}
with a double-well potential that has two physical parameters, $U_0$ and $d$, representing
the potential's depth and width.  This problem has been the subject of several recent studies  \cite{downing2013,zhang2014,agbool2014,dong2015,roy2015}. Besides being a useful model for a wide variety of applications, from heterostructure physics to the trapping of Bose-Einstein condensates,  it becomes an algebraically solvable system when certain constraints on the potential parameters $U_0$ and $d$ are satisfied. Under suitable transformations of the dependent and independent variables, the equation itself transforms into the poorly understood confluent Heun-type equation. The interesting spectral problem studied in this paper illuminates the contribution  of the confluent Heun equation to mathematical physics, and bridges an elusive physics problem to mathematical analysis.  In the present work, we introduce a concrete approach to find both analytic and approximate solutions for a class of hyperbolic potentials given by:
\begin{align}\label{eq2}
-\dfrac{\hbar}{2\mu}\dfrac{d^2\psi}{dx^2}+V_m(x;U_0,d)\psi=E\psi,\qquad V_m(x;U_0,d)=-U_0\dfrac{\sinh^{2m}(x/d)}{\cosh^{2m+2}(x/d)},\qquad \psi(\pm \infty)=0,\quad m=0,1,2,\dots.
\end{align}
This potential family includes, for $m=0$, the classical modified P\"oschl-Teller potential $V(x)=-U_0/\cosh^2(x)$, one of the few exactly solvable potentials in quantum mechanics \cite{rosen,flugge}.
\section{A class of potentials}\label{hyperbolicPot}
\begin{figure}[ht]
\centering 
\includegraphics[width=5in, height=3in]{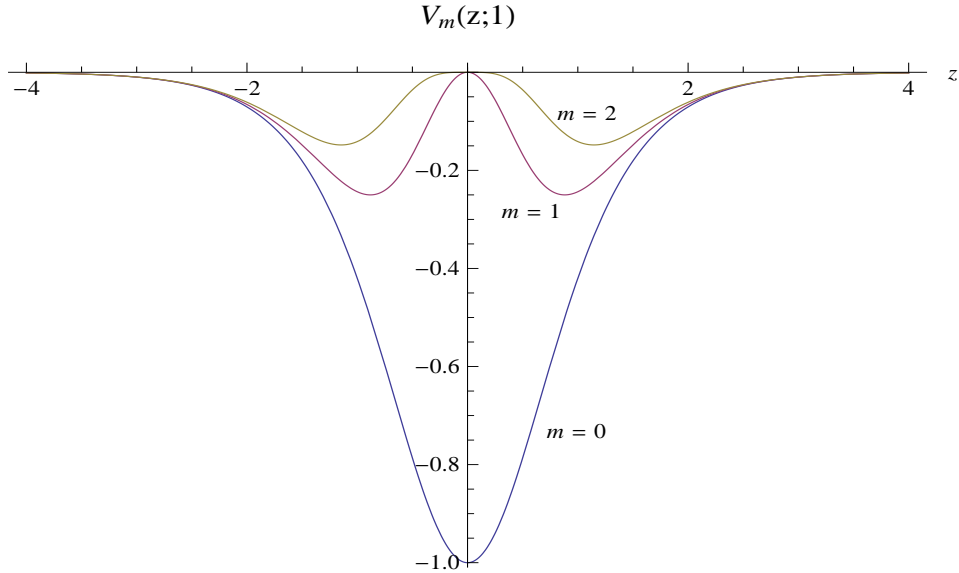}
\caption{The potential $V_m(z;1)=-{\sinh^{2m}(z)}/{\cosh^{2m+2}(z)}$ for $m=0,1,2$.}
\label{Fig1}
\end{figure}

\noindent Although the potential family $V(x;U_0,d)$ characterized by the two parameters $U_0$ and $d$, a simple change of variable $z=x/d$ transforms the equation into the following one-parameter Schr\"odinger equation 
 \begin{align}\label{eq3}
\left[-\dfrac{d^2}{dz^2}+V_m(z;v)\right]\psi=\varepsilon \psi,\quad V_m(z;v)=-v\dfrac{\sinh^{2m}(z)}{\cosh^{2m+2}(z)},\quad v>0,\quad -\infty<z<\infty,\quad \psi(\pm \infty)=0,
\end{align}
where $v={2\mu\, U_0\,d^2}/{\hbar^2}$ and $\varepsilon ={2\mu\,E\, d^2}/{\hbar^2}.$ The graph of the potential $V_m(z;1)$ for different values $m=0,1,2$ is  displayed in Fig. \ref{Fig1}. Clearly, $V_{m+1}(z)>V_{m}(z)$ for all $m$.
 
\noindent The minimum of the potential occurs at
$z=\pm (\cosh^{-1}(1+2m))/2,$
with a minimum value of $
V_{\rm min}(z;v)=-{m^m}/{(1+m)^{1+m}}v.$ 
We observe that $
\lim_{m\to \infty }V_{\rm min}(z,v)=0.$ Since
\begin{align}\label{eq4}
\int_{-\infty}^\infty V(z)dz&=
-v\int_{-\infty}^\infty \tanh^{2m}(z)\mbox{sech}^{2}(z) dz=-\frac{2v}{1+2 m}<0, \quad m=0,1,2,\dots.
\end{align}
the potential $V_m(z;v)$ has at least one negative eigenvalue \cite{martin} for any positive value of $v$ with eigenvalues $\epsilon$ satisfying $V_{\rm min}(z,v)<\varepsilon<0$. For each $m\ge 0$, the hyperbolic potential \eqref{eq3} has a finite number of bound-states $\mathfrak N$ with degeneracy one, and an upper bound on $\mathfrak N$ given \cite{chadan2003}  by:
\begin{align}\label{eq5}
\mathfrak N< 1+\sqrt{2}\left[
\int_{-\infty}^\infty z^2V(z)dz\int_{-\infty}^\infty V(z)dz\right]^{1/4}.
\end{align} 
However, for the hyperbolic potential \eqref{eq2}, 
 \begin{align*}
\int_{-\infty}^\infty z^2V(z)dz&= -\dfrac{4v}{2m+1}\left(\frac{\pi^2}{24}+\sum_{j=0}^{m-1}\frac{1}{4(m-j)}\left(\log(4)+H_{(2(m-j)-1)/2}\right)\right)\end{align*}
where $H_m$ is the $m^{th}$ Harmonic number $H_m=\sum_{k=1}^m 1/k$.
Thus, for each $m\ge 0$, the number of bound state energies is bounded above by 
\begin{align}\label{eq6}
\mathfrak N< 1+ \sqrt{\frac{v}{2m+1}} \left(\frac{4\pi^2}{3}+8\sum_{j=0}^{m-1}\frac{\log(4)+H_{(2m-2j-1)/2}}{(m-j)}\right)^{1/4},\quad m=0,1,2,\dots.
\end{align}
For example, for the modified P\"oschl-Teller potential, the number of the bound states is bounded above by $\mathfrak N<1+\sqrt{2\pi v}/3^{1/4}$.
\section{General differential equation}\label{DiffEq}
\noindent The change of variable $\eta=1/\cosh^2(z)$
maps the infinite interval $-\infty<z<\infty$ into $0<\eta \leq1$.  Since the potential $V(z)$ is an even function, the energy eigenfunctions may be classified as even $\psi_{+}(z)$ or odd $\psi_{-}(z)$ functions of $z$. If we write $\psi(z) = \phi(\eta),$ the boundary condition requirement $\psi(\pm \infty)=0$  is equivalent  to the condition $\phi(0)=0$. Thus, the mapping has the feature that the change of variable $z\rightarrow \eta$ covers the interval $(0,1)$ twice and vanishes at the end point $\eta=1$ only once corresponding to $z=0$. Thus, for both even and odd cases, for each zero of the wave function $\phi(\eta)$, where $\eta\in (0,1)$, there are two zeros of the wave function $\psi(z)$ for $z\in (-\infty,\infty)$; whereas, in the odd case, $\psi(z)$ has one extra zero $\psi_{-}(0) = \phi(1) = 0.$  This change of variable reduces equation \eqref{eq3} to
\begin{align}\label{eq7}
4\eta^2(1-\eta)\dfrac{d^2\phi(\eta)}{d\eta^2}+\left[4\eta-6\eta^2
\right]\frac{d\phi(\eta)}{d\eta}+\left(\varepsilon +v\,\eta\,(1-\eta)^{m}\right)\phi(\eta)=0,
\end{align}
with boundary condition(s) $\phi(0)=0$ and $\phi(1)\neq 0$ or $\phi(0)=0$ and $\phi(1)=0$.
For each $m\ge 0$, the differential equation \eqref{eq7} has two regular singular points at $\eta=0$ with exponents $\{\pm \sqrt{-\varepsilon}/2\}$ where $\varepsilon <0$ and at the singular point $\eta=1$ with exponents $\{0,1/2\}$. For the singular point at $\eta =\infty$,  the transformation $\xi=1/{\eta}$ is used
and the resulting equation is examined for the regularity at $\xi=0$. It is not difficult to deduce that the resulting equation has a regular singular point at $\xi=0$ with exponents $\{(1\pm\sqrt{1+8v})/4\}$ if $m=0$. On other hand, the transformed equation has an irregular singular point at $\xi=0$ for all $m\ge 1$. Consequently, the general solution of \eqref{eq7} may assume the form
\begin{align}\label{eq8}
\phi(\eta)=\eta^\alpha (1-\eta)^\beta e^{-\gamma\,\eta} f(\eta),\quad \eta\in (0,1)
\end{align}
where $\alpha=\sqrt{-\varepsilon}/2$ and $\beta$ takes either the value of $\beta=0$ or $\beta=1/2$. The parameter $\gamma=0$ for $m=0$ is  used to sustain the regularity at infinity, and $\gamma\geq 0$ for $m\geq 1$. Again, because of the two possible values of the parameter $\beta$, we have to distinguished between two cases: For $\beta=0$, the wave function \eqref{eq8} vanishes at $\eta=0$ and since $\eta\neq 1$, the boundary condition $\phi(0)=0$ equivalent to $\psi(\pm \infty)=0$ and the resulting wave function \eqref{eq8} is even $\psi_{+}$. For $\beta=1/2$, the wave function \eqref{eq8} vanishes at $\eta=0$ in addition to $\eta= 1$, the boundary conditions $\phi(0)=\phi(1)=0$, in this case, equivalent to $\psi(\pm \infty)=\psi(0)=0$ and the resulting wave function $\psi_{-}(z)$ is odd with respect to $z$. On substituting \eqref{eq8} into \eqref{eq7}, the unknown function $f(\eta)$ has the following differential equation
\begin{align}\label{eq9}
f''(\eta)&+\left(\dfrac{2\alpha+1}{\eta}+\dfrac{4\beta+1}{2(\eta-1)}-2\gamma\right)f'(\eta)+\bigg(\dfrac{\beta(2\beta-1)}{2(\eta-1)^2}+\dfrac{4\alpha^2+\varepsilon}{4\eta^2}+\gamma^2+\frac{2\alpha+4\beta+8 \alpha\beta-\varepsilon-2\gamma-8\beta\gamma}{4 (\eta-1)}\notag\\
&+\frac{\varepsilon-2\alpha-4\beta-8 \alpha\beta-4\gamma-8\beta\gamma}{4\eta}+\dfrac{v}{4\eta}(1-\eta)^{m-1}\bigg)f(\eta)=0.
\end{align}
Since $\alpha=\sqrt{-\varepsilon}/2$ for either value of $\beta=0$ or $\beta=1/2$, the term $\beta(2\beta-1)=0$ and the equation \eqref{eq9} reduce to
\begin{align}\label{eq10}
f''(\eta)+\left(\dfrac{2\alpha+1}{\eta}+\dfrac{4\beta+1}{2(\eta-1)}-2\gamma\right)f'(\eta)&+\bigg(\gamma^2+\frac{2\alpha+4\beta+8 \alpha\beta-\varepsilon-2\gamma-8\beta\gamma}{4 (\eta-1)}\notag\\
&+\frac{\varepsilon-2\alpha-4\beta-8 \alpha\beta-4\gamma-8\beta\gamma}{4\eta}+\dfrac{v}{4\eta}(1-\eta)^{m-1}\bigg)f(\eta)=0.
\end{align}
This is the general differential equation that we attempt to solve, either analytically or approximately, for the non-negative integer $m=0,1,2,\dots$.
\section{The modified P\"oschl-Teller potential}
\noindent For $m=0$, the potential $V_{m=0}(z)=-v/\cosh^2(z)$ is the classical modified P\"oschl-Teller potential often used as a realistic model for molecular potentials. The one-dimensional Schr\"odinger equation with this  potential has been analyzed long ago \cite{rosen,teller} and has been studied extensively ever since. Thus, we briefly outline our solution within the application of the general equation \eqref{eq10}. For $m=0$, $\gamma=0$ and the equation \eqref{eq10} reduces to
\begin{align}\label{eq11}
f''(\eta) &+\left(\dfrac{(3+4\alpha+4\beta) \eta-2-4 \alpha }{2\eta(\eta-1)}\right)f'(\eta)+\bigg(\frac{4\beta+2\alpha (1+4\beta)- \varepsilon-v}{4\eta (\eta-1)}\bigg)f(\eta) =0,\quad 0<\eta<1.
\end{align}
Equation \eqref{eq11} has three regular singular points at $\eta=0,1,\infty$, and according the general theory of the hypergeometric equation \cite{kita}, the general solution is expressible in terms of Gauss's hypergeometric function ${}_2F_1(\alpha,\beta;\gamma;z)=\sum_{k=0}^\infty (\alpha)_k(\beta)_k/[(\gamma)_k\, k!]z^k,$ where $(z)_k$ is the Pochhammer symbol defined in terms of Gamma function by $(z)_k=z(z+1)\dots(z+k-1)=\Gamma(z+k)/\Gamma(z)$. Indeed, it is not difficult to show that the differential equation has the solution, see also \cite{saad2014},
\begin{align}\label{eq12}
f(\eta)&=\, _2F_1\left(\beta+\frac{1+2\sqrt{-\varepsilon}-\sqrt{1+4v}}{4},\beta+\frac{1+2\sqrt{-\varepsilon}+\sqrt{1+4v}}{4};1+2\alpha;\eta\right),
\end{align}
where $\beta(2\beta-1)=0$ and $4\alpha^2+\varepsilon=0$ have been employed.  
The infinite series representation of \eqref{eq12} terminates to an $n$-degree polynomial if
 \begin{align}\label{eq13}
  \beta+\frac{1+2\sqrt{-\varepsilon}-\sqrt{1+4v}}{4}=-n,\qquad n=0,1,2,\dots,
  \end{align}
  that yields $\sqrt{-\varepsilon}=(-1-4n-4\beta+\sqrt{1+4v})/2$ . Since the left-hand side of this equation is positive, it is necessary that $-1-4n-4\beta+\sqrt{1+4v}>0$ which bounds on the number of the eigenenergies given by the formula $
 n<\left(-1-4\beta+\sqrt{1+4v}\right)/4.
$
Thus, in summary, the exact solutions of Schr\"odinger's equation 
\begin{equation}\label{eq14}
\left[-\dfrac{d^2}{dz^2}-\dfrac{v}{\cosh^{2}(z)}\right]\psi_n=\varepsilon_n \psi_n,\qquad -\infty<z<\infty,\qquad \psi_n(\pm \infty)=0,
\end{equation}
are given explicitly by
\begin{align}\label{eq15}
\psi_n(z)&=\left\{ \begin{array}{ll}
{\mbox{sech}^{\sqrt{-\varepsilon_n}}\left(z\right)}\,
{}_2F_1\left(-n,\frac{1}{2}+\sqrt{-\varepsilon_n}+n;1+\sqrt{-\varepsilon_n};\mbox{sech}^2\left(z\right)\right), &\mbox{ (Even states, $\beta=0$, $n=0,1,\dots$),} \notag\\ \notag\\
{\mbox{sech}^{\sqrt{-\varepsilon_n}}\left(z\right)}\, {\tanh\left(z\right)}\,
{}_2F_1\left(-n,\frac{3}{2}+\sqrt{-\varepsilon_n}+n;1+\sqrt{-\varepsilon_n};\mbox{sech}^2\left(z\right)\right),&\mbox{ (Odd states, $\beta=\frac12$, $n=0,1,\dots$).}
       \end{array} \right.\\
\end{align}
where
\begin{align}\label{eq16}
\varepsilon_n=-\frac{1}{4} \left(-1-4\beta-4n+\sqrt{1+4v}\right)^2,
\end{align}
 for $n< \left(-1-4\beta+\sqrt{1+4 v}\right)/4$ or $v>2(\beta+n)(1+2\beta+2n)$ where $ \beta=0,1/2$. 
 The exact number of the bound-states of the modified P\"oschl-Teller potential, given $v$ and $\beta$, is precisely
 \begin{align}\label{eq17}
 \mathfrak N=1+\left\lfloor\dfrac{-1-4\beta+\sqrt{1+4 v}}{4}\right\rfloor,
  \end{align}
   where $\lfloor x\rfloor$ is the greatest integer less than or equal to $x$.  
\section{The Asymptotic iteration method}
\noindent The asymptotic iteration method (AIM) is an iterative algorithm originally introduced  \cite{hakan2003} to investigate the analytic
and approximate solutions of a second-order linear differential equation
\begin{equation}\label{eq18}
y''=\lambda_0(r) y'+s_0(r) y,\quad\quad ({}^\prime={d\over dr})
\end{equation}
where $\lambda\equiv \lambda_0(r)$ and $s_0\equiv s_0(r)$ are $C^{\infty}(a,b)-$differentiable
functions. AIM states \cite{hakan2003}: \emph{Given $\lambda_0$ and $s_0$ in
$C^{\infty}(a,b),$ the differential equation \eqref{eq18} has the
general solution
\begin{align}\label{eq19}
y(r)&= \exp\left(-\int\limits^{r}{s_{n-1}(t)\over \lambda_{n-1}(t)} dt\right) \left[C_2
+C_1\int\limits^{r}\exp\left(\int\limits^{t}\left[\lambda_0(\tau) +
{2s_{n-1}\over \lambda_{n-1}}(\tau)\right] d\tau \right)dt\right]
\end{align}
where $C_1$ and $C_2$ are the integration constants, if for sufficiently large $n>0$
\begin{equation}\label{eq20}
\delta_n=\lambda_n s_{n-1}-\lambda_{n-1}s_n=0.
\end{equation}
The AIM sequences $\lambda_n$ and $s_n$, $n=1,2,\dots$, are computed recursively using
\begin{equation}\label{eq21}
\lambda_{n}=
\lambda_{n-1}^\prime+s_{n-1}+\lambda_0\lambda_{n-1}\hbox{ ~~and~~
} s_{n}=s_{n-1}^\prime+s_0\lambda_{n-1}.
\end{equation}}

\noindent Over the past decade, AIM has proved to be an efficient and effective algorithm for solving many eigenvalue problems that occur in relativistic and non-relativistic quantum mechanics. The first step in applying the AIM algorithm is to construct a product of an asymptotic solution to the given boundary-value problem with an unknown function (to be determine by AIM). Thus the original problem is transformed into the eigenvalue problem with the  form \eqref{eq18}. The second step is to evaluate the termination condition \eqref{eq20} by using the AIM sequences $\{\lambda_n(r)\}$ and $\{s_n(r)\}$  recursively, as given by \eqref{eq21}. The resulting expressions for $\delta_n$ are usually functions of the (unknown) eigenvalue $E$ and the independent variable $r$. A one-dimensional root-finding method is then employed to evaluate the roots of the equation $\delta_n(E,r)=0$ for a suitable  initial value $r_0$  of $r$. If the eigenvalue problem is solvable, the number of wave function zeros is equal to the iteration number $n$ and the roots of the termination condition \eq{eq19} are precisely the exact eigenvalues $\varepsilon_n$ regardless of the given initial value  $r_0$. In other cases, the iteration sequence is arbitrarily stopped and AIM is employed as an approximation method with the advantage of being a simple programmable algorithm. The number of iterations $N$, in this case, does not generally correspond to the energy level, but, as the iteration number $N$ increases, the sequence of roots of the termination condition converges to the desired eigenvalue. The rate of convergence, however, depends mainly on the asymptotic solution initially employed and on a suitable initial value $r_0$ of $r$ \cite{brodie}. There is as yet no reliable general criterion to determine such a value.    AIM users currently apply various approaches to choose $r_0$, such as the position of the deepest point of the potential (if it is not zero),  the location of the maximum of an approximate ground-state wave function, or the centre of the bounded region.  If the domain of the problem is initially unbounded, it helps first to transform it to a  bounded domain, as we have done for the present application.  Fortunately the quality of the eigenvalue approximations is usually found to be stable with respect to choices of $r_0$. 

 \vskip0.1true in
\noindent The modified P\"oschl-Teller potential serves as a perfect test example to examine the accuracy of the AIM algorithm and as a benchmark for the more difficult problems to be solved. In Table \ref{Tab1}, we check our computer program, written using Maple 16 running on a MacBook Pro 2.5 GHz Intel Core 17 with 8GB RAM against the exact values as given by formula \eqref{eq16}. The number of the bound-states indicated by AIM is in complete agreement with the exact number as given by \eqref{eq17}.
\begin{center}
\begin{table}[htbp]
  \begin{tabular}{|c|c|c|c|c|}
    \hline
    $\beta$&$n$&$v$& $\varepsilon_n\, (Exact)$&$\varepsilon_n\, (AIM)$\\
        \hline
        $0$&$0$&$1$& $~-0.381~966~011~250~105~151~79$&$ ~~~-0.381~966~011~250~105~151~79_{(3;0.627)}$\\
        \hline      
        & $0$&$4$&  $~-2.438~447~187~191~169~725~09$&$~~-2.438~447~187~191~169~725~09_{(3;0.052)}$\\  \hline      
        & $0$ &$9$&  $~-6.458~618~734~850~890~155~50$&$~~-6.458~618~734~850~890~155~50_{(3;0.052)}$\\  
        & $1$&& $~-0.293~093~674~254~450~777~50$&$~~-0.293~093~674~254~450~777~50_{(3;0.229)}$\\  \hline
        & $0$ &$16$& $~-12.468~871~125~850~725~173~82$&$~-12.468~871~125~850~725~173~82_{(3;0.053)}$\\  
        & $1$&~&  $~-~2.344~355~629~253~625~869~08$&$~~-2.344~355~629~253~625~869~08_{(3;0.416)}$\\  
              \hline
        & $0$ &$25$&  $~-20.475~062~189~439~554~864~89$&$~-20.475~062~189~439~554~864~89_{(3;0.115)}$\\  
        & $1$&~&  $~~-6.375~310~947~197~774~324~45$&$~-~6.375~310~947~197~774~324~45_{(3;0.059)}$\\  
        & $2$&~&  $~~~-0.275~559~704~955~993~784~01$&$~~-~0.275~559~704~955~993~784~01_{(4;0.375)}$\\  
              \hline
        \end{tabular}
        \caption{Exact eigenenergies $\beta=0$ evaluated using formula \eqref{eq16} comparing with AIM results. Number of iterations along with the computational  times in seconds, used by AIM, are given as subscripts  of the column $\varepsilon_{n}(AIM)$. In all of our computations, the iterative process started with the value $r_0=1/2$.}
\label{Tab1}
\end{table}
\end{center}

\section{A class of hyperbolic double-well potentials}
\noindent In this section, the case $m=1$ is analyzed, namely  Schr\"odinger's equation 
\begin{equation}\label{eq22}
-\dfrac{d^2\psi}{dz^2}-v\,\dfrac{\sinh^2(z)}{\cosh^4(z)}\psi=\varepsilon\,\psi,\qquad -\infty <z<\infty,\quad \psi(\pm\infty)=0.
\end{equation}
For this equation, the assumed general solution \eqref{eq8} leads in general to the differential equation \eqref{eq10}, that is to say
\begin{align}\label{eq23}
\eta^2(4-4\eta)\dfrac{d^2\phi(\eta)}{d\eta^2}+\eta\left[4-6\eta
\right]\frac{d\phi(\eta)}{d\eta}+\left(\varepsilon + v\, \eta-v\,\eta^2\right)\phi(\eta)=0,
\end{align}
and we find in this case
\begin{align}\label{eq24}
f''(\eta)&+\left(\dfrac{2\alpha+1}{\eta}+\dfrac{4\beta+1}{2(\eta-1)}-2\gamma\right)f'(\eta)\notag\\
&+\bigg(\gamma^2+\frac{2\alpha+4\beta+8 \alpha\beta-\varepsilon-2\gamma-8\beta\gamma}{4 (\eta-1)}+\frac{\varepsilon+v-2\alpha-4\beta-8 \alpha\beta-4\gamma-8\beta\gamma}{4\eta}\bigg)f(\eta)=0,
\end{align}
where $\beta(2\beta-1)=0$ and $4\alpha^2+ \varepsilon=0$ has been used. Equation \eqref{eq24} does not admit any polynomial solution \cite{saad2008}, as the criterion for polynomial solutions is not satisfied. Further, since $\eta=\infty$  is an irregular singular point for arbitrary value  of $\gamma\geq 0$, we set $\gamma=0$
and this reduces equation \eqref{eq24} to
\begin{align}\label{eq25}
f''(\eta)&=-\left(\frac{1+4\beta}{2 (\eta-1)}+\frac{1+\sqrt{-\varepsilon}}{\eta}\right)f'(\eta)-\left(\frac{v\eta+\sqrt{-\varepsilon} + 4 \beta + 4\beta\sqrt{-\varepsilon} - \varepsilon -v}{4\eta(\eta-1)}
\right)f(\eta).
\end{align}
For this differential equation, the coefficients of the infinite series solution 
\begin{align}\label{eq26}
f(\eta)=\sum_{n=0}^\infty c_n \eta^n,
\end{align}
by Frobenius's  method 
obey the recurrence relation
\begin{align}\label{eq27}
c_0&=1,\quad c_1=\frac{4 \beta+2 \alpha (1+4\beta)-\varepsilon-v}{4+8\alpha},\notag\\
c_n&+\dfrac{\left(v+4 \beta+\varepsilon+\alpha (6-8\beta-8 n)+6 n-8\beta n-4 n^2-2\right)}{4 n (2\alpha+n)} c_{n-1}-vc_{n-2}=0,\quad n\geq 2.
\end{align}
These coefficients have an interesting property that allows us to evaluate the series coefficients of \eqref{eq26} as
\begin{align}\label{eq28}
f(\eta)=\sum_{n=0}^\infty \dfrac{P_n(\alpha)}{n! \,(1+2\alpha)_n} \left(\dfrac{\eta}{4}\right)^n,
\end{align}
where for $\beta=0$, the polynomials $\{P_n(\alpha)\}_{n=0}^\infty$ satisfy the three-term recurrence relation 
\begin{align}\label{eq29}
P_{n+1}(\alpha)=(2n(2n+1)+(8n+2)\alpha+4\alpha^2-v)P_n(\alpha)+4n\,v\,(n+2\alpha)P_{n-1}(\alpha),\quad P_{-1}(\alpha)=0,P_0(\alpha)=1,
\end{align}
while for $\beta=1/2$, the series solution \eqref{eq26} takes the form
\begin{align}\label{eq30}
f(\eta)=\sum_{n=0}^\infty \dfrac{{\mathcal P}_n(\alpha)}{n!\, (1+2\alpha)_n} \left(\dfrac{\eta}{4}\right)^n,
\end{align}
where now  the polynomials $\{\mathcal P_n(\alpha)\}_{n=0}^\infty$ satisfy the recurrence relation 
\begin{align}\label{eq31}
\mathcal P_{n+1}(\alpha)=(2(n+1)(2n+1)+(8n+6)\alpha+4\alpha^2-v)\mathcal P_n(\alpha)+4n\,v\,(n+2\alpha)\mathcal P_{n-1}(\alpha),\quad P_{-1}(\alpha)=0,P_0(\alpha)=1.
\end{align}
The discrete spectrum of the Hamiltonian \eqref{eq21} evaluated using AIM initiated with
\begin{align}\label{eq32}
\lambda_0=-\left(\frac{1+4\beta}{2 (\eta-1)}+\frac{1+\sqrt{-\varepsilon}}{\eta}\right),\quad and\quad s_0=-\left(\frac{v\eta+\sqrt{-\varepsilon} + 4 \beta + 4\beta\sqrt{-\varepsilon} - \varepsilon -v}{4\eta(\eta-1)}
\right)
\end{align}
are reported in Tables \ref{Tab2} and \ref{Tab3}. Starting with $r_0=1/2\in (0,1)$, in Table \ref{Tab2} we report our finding of $\varepsilon$ using the roots of the termination condition \eqref{eq19} determined accurately to the first 24 decimal places, along with the number of iteration $N$ used by AIM. In Table \ref{Tab3}, we also report the eigenvalues for higher values of the parameter $v$. AIM converges fast as indicated by the low number of iterations for very high precision of the eigenvalues for the given potential strength $v$. With this finding, the coefficients of the wave function are easily computed using Eqs. \eqref{eq29} and \eqref{eq31}. 

\begin{table}[htbp]
  \begin{tabular}{|c|c|c|c|c|c|}
    \hline
    $\beta$&$n$&$v$&$V_{min}$&$\varepsilon_n(AIM)$&$N_{iteration}$\\
        \hline
        $0$&$0$&$0.0001$&$-0.000~025$& $-0.000~000~001~110~997~544~530~833_{(0.173)}$&$5$\\
        \hline    
        $0$&$0$&$0.0004$&$-0.000~100 $&$-0.000~000~017~770~512~138~752~699_{(0.767)}$&$6$\\ 
                      \hline    
 $0$&$0$&$0.0009$&$-0.000~225$&$-0.000~000~089~917~289~559~488~488_{(0.545)}$&$5$\\ 
                      \hline 
 $0$&$0$&$0.0016$&$-0.000~400 $&$-0.000~000~283~980~113~514~486~522_{(0.687)}$&$6$\\ 
                      \hline             
         $0$&$0$&$0.0025$&$-0.000~625$&$-0.000~000~692~675~074~023~441~258_{(0.741)}$&$6$\\ 
                      \hline             
                      \end{tabular}
                      \caption{The eigenvalues for the potential $-v\sinh^{2}(z)/\cosh^{4}(z)$ for very small values of the potential parameter $v$. The computational times in seconds, used by AIM, are given as subscript values of the column $\varepsilon_n(AIM)$.}
                      \label{Tab2}
\end{table}

\begin{table}[!htbp]
  \begin{tabular}{|c|c|c|c|c|c|}
    \hline
    $\beta$&$n$&$v$&$V_{min}$&$\varepsilon_n(AIM)$&$N_{iteration}$\\
        \hline
        $0$&$0$&$5$&$-1.25$& $-0.547~952~205~095~460~959~101~243_{( 1.558)}$&$14$\\
        $1/2$&$0$&~&~&$-0.069~381~268~987~066~025~562~792_{(1.323)}$&$13$\\
        \hline    
        $0$&$0$&$10$&$-2.5$& $-1.284~258~416~184~695~724~376~712_{( 2.848)}$&$16$\\
        $1/2$&$0$&~&~&$-0.625~853~590~393~309~267~849~407_{(2.072)}$&$14$\\
        \hline       
        $0$&$0$&$20$&$-5$& $-2.912~882~550~184~603~884~378~849_{(4.999)}$&$18$\\
        $1/2$&$0$&~&~&$-2.249~697~456~457~806~280~295~288_{(3.678)}$&$16$\\
            $0$&$1$&$~$&$~$& $-0.048~478~263~588~559~450~977~637_{(4.936)}$&$18$\\       
              \hline       
        $0$&$0$&$30$&$-7.5$& $-4.674~864~616~067~671~875~486~965_{(6.830)}$&$19$\\
        $1/2$&$0$&~&~&$-4.103~428~902~050~365~691~474~461_{(5.750)}$&$18$\\
            $0$&$1$&$~$&$~$& $-0.578~712~111~337~799~069~384~936_{(8.585)}$&$20$\\  
           $1/2$&$1$&~&~&$-0.082~558~916~307~354~686~674~354_{(6.126)}$&$18$\\      \hline       
        $0$&$0$&$50$&$-12.5$& $-8.462~774~605~628~490~576~718~186_{(12.437)}$&$21$\\
        $1/2$&$0$&~&~&$-8.074~319~923~536~337~449~665~977_{(10.623)}$&$20$\\
            $0$&$1$&$~$&$~$& $-2.383~275~941~780~201~266~826~343_{(17.036)}$&$22$\\  
           $1/2$&$1$&~&~&$-1.481~421~225~325~608~520~260~491_{(6.119)}$&$20$\\      
              \hline       
        $0$&$0$&$100$&$-25$& $-18.764~147~649~169~376~439~784~972_{(24.872)}$&$24$\\
        $1/2$&$0$&~&~&$-18.616~900~710~859~325~956~850~023_{(15.931)}$&$22$\\
            $0$&$1$&$~$&$~$& $-~8.595~652~745~858~308~209~944~525_{(34.286)}$&$25$\\  
           $1/2$&$1$&~&~&$-~7.839~606~197~742~417~788~821~255_{(26.372)}$&$24$\\    
            $0$&$2$&$~$&$~$& $-~2.184~901~373~795~703~192~078~575_{(24.656)}$&$25$\\    
            $1/2$&$2$&~&~&$-~1.208~534~073~121~443~535~689~563_{(21.112)}$&$24$\\                                            \hline                       
            \end{tabular}
             \caption{The eigenvalues for the potential $-v\sinh^{2}(z)/\cosh^{4}(z)$ for higher values of the parameter $v$. The computational times in seconds, used by AIM, are given as subscript values of the column $\varepsilon_n(AIM)$}
                      \label{Tab3}\end{table}

\section{Another class of hyperbolic double-well potential}
\noindent In this section, the case $m=2$ is examined and both the quasi-exact and the approximate solutions for the entire discrete spectrum are evaluated for the Schr\"odinger equation 
\begin{equation}\label{eq33}
-\dfrac{d^2\psi}{dz^2}-v\,\dfrac{\sinh^4(z)}{\cosh^6(z)}\psi=\varepsilon\,\psi,\qquad -\infty<z<\infty,\quad \psi(\pm \infty)=0.
\end{equation}
For this equation, the assumed solution \eqref{eq8} of the differential equation \eqref{eq10},
\begin{align}\label{eq34}
4\eta^2(1-\eta)\dfrac{d^2\phi(\eta)}{d\eta^2}+\left[4\eta-6\eta^2
\right]\frac{d\phi(\eta)}{d\eta}+\left(\varepsilon +v\,\eta\,(1-\eta)^{2}\right)\phi(\eta)=0, \quad \phi(0)=\phi(1)=0,
\end{align}
becomes explicitly 
\begin{align}\label{eq35}
(2\eta^2-2\eta)f''(\eta)&+(-4\gamma \eta^2+(3 + 4 \alpha + 4 \beta + 4 \gamma)\eta-2(1+2\alpha))f'(\eta)\notag\\
&+(\gamma(2\gamma -3-
4\alpha-4\beta)\eta + \alpha+2 \alpha^2
+2\beta+4\alpha\beta+2 \gamma+4\alpha\gamma-2\gamma^2)f(\eta)=0
\end{align}
where we have used the relations $2\beta^2-\beta=0,~4\alpha^2+\varepsilon=0,$ and $4\gamma^2-v=0
$. As noticed earlier \cite{downing2013}, this is Heun's confluent-type differential equation \cite{ronveaux}.
It has a solution around the regular singular point $\eta=
0$ given in terms of the confluent Heun function \cite{downing2013} that can be explicitly expressed using, for example,  Maple computing software. However, we introduce in Theorem \ref{Thm1} a slightly easier method for evaluating the exact solutions in terms of a recurrence relation instead of the correlation between polynomial equations and matrix determinants usually used \cite{ronveaux}.  We first give a general result valid for a class of differential equations.
\begin{theorem}\label{Thm1}
The necessary condition for the existence of $N$-degree polynomial solutions of the differential equation
\begin{align}\label{eq36}
(a_2z^2+a_1z)f''(z)+(b_2z^2+b_1z+b_0)f'(z)-(\tau_1\, z+\tau_0)f(z)=0
\end{align}
is \begin{align}\label{eq37}
\tau_1=N b_2,\qquad N=0,1,2,\dots.
\end{align}
and the polynomial solutions are give explicitly by
\begin{align}\label{eq38}
f_N(z)=\sum_{k=0}^N \dfrac{P_k(\tau_0)}{k!\,a_1^k\left({b_0}/{a_1}\right)_k}\,z^k,
\end{align}
where, for each $N$, the finite sequence of the polynomials $\{P_k(\tau_0)\}_{k=0}^N$ satisfies a three-term recurrence relation, for $0\leq k\leq n+1$,
\begin{align}\label{eq39}
P_{k+1}(\tau_0)=(\tau_0-k(k-1)a_2-kb_1)P_k(\tau_0)+kb_2(N-k+1)((k-1)a_1+b_0)P_{k-1}(\tau_0),
\end{align}
initialized with $P_{-1}(\tau_0)=0,~P_0(\tau_0)=1$. 
 \end{theorem}

\noindent The proof of this theorem is given in the appendix along the explicit forms of the first few polynomial solutions.

\noindent Direct comparison of equation \eqref{eq35} with \eqref{eq36} gives the necessary condition for polynomial solutions of \eqref{eq34} as
\begin{align}\label{eq40}
3 + 4 \beta + 2 \sqrt{-\varepsilon} + 4 N - \sqrt{v}=0
\end{align}
from which we obtain for the following formula for exact eigenvalues
\begin{align}\label{eq41}
\varepsilon_N=-\frac{1}{4} \left(\sqrt{v}-3-4\beta-4 N\right)^2,\qquad v>(3+4\beta+4N)^2,\qquad \beta=0,1/2.
\end{align}
It should be clear that $N$ is the degree of polynomial solution, not necessary the number of nodes $n$ of the full wave function, as discussed earlier. For each $N$, the polynomial solution is  given by
\begin{align}\label{eq42}
f_N(\eta)=\sum_{k=0}^N \dfrac{P_k}{k!\,\left(1 + \sqrt{-\varepsilon}\right)_k}\,\left(-\dfrac{\eta}{2}\right)^k,
\end{align}
where the polynomial coefficients $\{P_k\}_{k,0}^N$ are evaluated in terms of $\varepsilon$ and $v$ using 
\begin{align}\label{eq43}
P_{k+1}&=\left(\frac{\varepsilon}{2}-\frac{\sqrt{-\varepsilon}}2(1+4\beta+4k+2\sqrt{v})+\frac{v}{2}-(2\beta+k) (1+2 k)-(1+2 k) \sqrt{v}\right)P_k\notag\\
&+4 k (N-k+1) \sqrt{v}\left(\sqrt{-\varepsilon}+k\right) P_{k-1},\qquad 0\leq k\leq N+1,
\end{align}
initialized by $P_{-1}=0$ and $P_0=1$. The sufficient condition for the polynomial solution is given explicitly by \eqref{eq43} for $k=N+1$. In the next subsections, the polynomial solutions of degree $N=0,1,2$, are discussed in detail. Higher order polynomial solutions may be constructed similarly.

\subsection{Zero-degree polynomial solution} 
\noindent We note that although only $\psi(z)$ are even or odd functions, the corresponding $\phi(\eta)$ functions
will be written with the same symmetry subscripts: thus $\psi_{\pm}(z)\longleftrightarrow\phi_{\pm}(\eta).$
\smallskip\hfil\break
\noindent In the case $N=0$, the constant solution
\begin{align}\label{eq44}
f_0(\eta)&=1,
\end{align}
is subject to the following two conditions, relating $\varepsilon$ and $v$, 
\begin{align}\label{eq45}
3+4\beta+2 \sqrt{-\varepsilon}-\sqrt{v}&=0,\qquad
\varepsilon-(4\beta+2\sqrt{v}+1)\sqrt{-\varepsilon}-4\beta-2\sqrt{v}+v=0.
\end{align}
The non-zero solutions of this system yields, for $\beta=0$,
$v=29+8\sqrt{13},$ and $\varepsilon =-(7+\sqrt{13})/2$
with wave function $\phi(\eta)$ given by 
 \begin{align*}
 \phi_{+}(\eta)=\eta^{\left(\sqrt{(7+\sqrt{13})/2}\right)/2} e^{-\sqrt{29+8\sqrt{13}}\, \eta/2},\qquad \phi_{+}(0)=0,\quad \phi_{+}(1)\neq 0,
 \end{align*}
 while for $\beta=1/2$, the polynomial solution \eqref{eq44} is subject to the constraints 
$v=125+16\sqrt{61}$ and  $\varepsilon=-(35+3\sqrt{61})/2,$
with the odd wave function 
\begin{align*}
\phi_{-}(\eta)=\eta^{\left(\sqrt{(35+3\sqrt{61})/2}\right)/2}\, \sqrt{1-\eta}\, e^{-\sqrt{125+16\sqrt{61}}\, \eta/2},\quad \phi_{-}(0)=\phi_{-}(1)=0.
\end{align*}
The corresponding full wave functions of Schr\"odinger's equation \eqref{eq32} are then
\begin{align}
\psi_{+}(z)&=\mbox{sech}^{(1+\sqrt{13})/2}(z) e^{-(4+\sqrt{13})\, \mbox{sech}^2(z)/2},\label{eq46}\\
\psi_{-}(z)&=\tanh(z)\,  \mbox{sech}^{{(3+\sqrt{61})/2}}(z)\, e^{-(8+\sqrt{61})\, \mbox{sech}^2(z)/2}.
\label{eq47}
\end{align}

\begin{figure}[!h]
\centering 
\includegraphics[width=3in, height=2.1in]{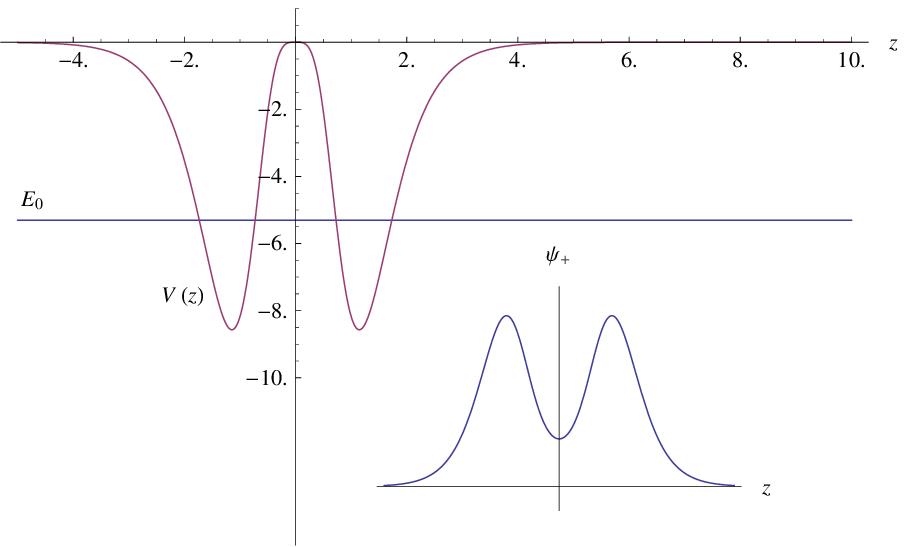}\qquad
\includegraphics[width=3in, height=2in]{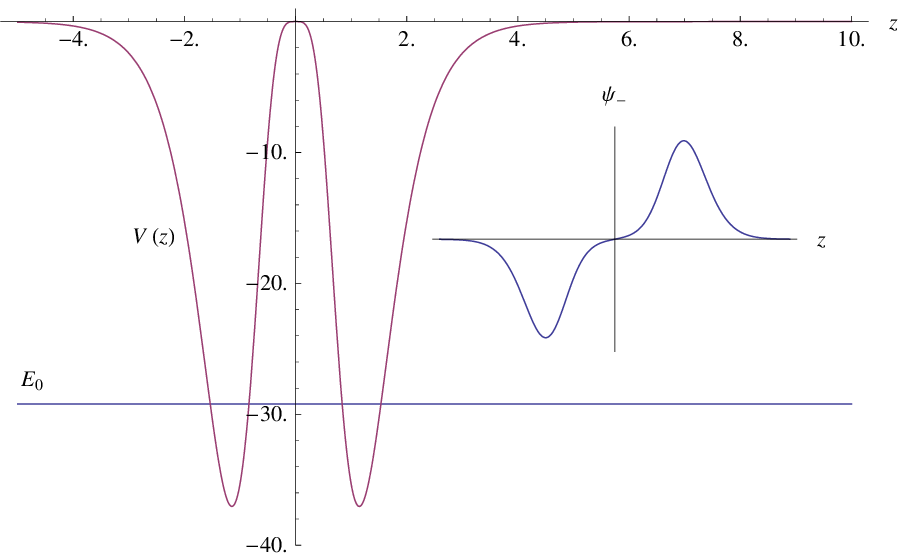}
\caption{Left: Plot of the potential $V(z)=-(29+8\sqrt{13})\sinh^4(z)/\cosh^6(z)$  along with the exact eigenvalue $E_0$ and the exact wave function $\psi_{sy}(z)$ (Inset). Right: Plot of the potential $V(z)=-(125+3\sqrt{61})\sinh^4(z)/\cosh^6(z)$  along with the exact eigenvalue $E_0$ and the exact wave function $\psi_{asy}(z)$ (Inset).}
\label{Fig2}
\end{figure}

\noindent The graph of these \emph{exact} bound-state wave functions are displayed in Fig. \ref{Fig2} along with the plot of the associated potential and the exact eigenvalue. We note that the minimum of the potential $V(x)=-v\,\sinh^4(z)/\cosh^6(z)$ is $V_{min}=-4v/27$,  if the potential strength  is fixed at $v=29+8\sqrt{13}$ and at $v=125+16\sqrt{61}$, the minimum of the potential is, respectively, $V_{min}= -4(29+8\sqrt{13})/27\sim -8.569~5\dots$ and $V_{min}=-4(125+16\sqrt{61})/27= -37.0317$. It is natural to ask whether the potential supports the existence of other bound-states beside the exact $\psi_{+}(z)$ and $\psi_{-}(z)$. To find out, we rely on AIM to evaluate all the possible (discrete) eigenvalues, including the exact ones, as a test example. Writing Eq. \eqref{eq34} as
$
f''(\eta)=\lambda_0(\eta) f'(\eta)+s_0(\eta) f(\eta),
$
where
\begin{align}\label{eq48}
\lambda_0(\eta)&=-\left(\dfrac{1+\sqrt{-\varepsilon}}{\eta}+\dfrac{4\beta+1}{2(\eta-1)}-\sqrt{v}\right),\notag\\
s_0(\eta)&=-\bigg(\frac{-4\beta\sqrt{v}+4\beta\sqrt{-\varepsilon}-\sqrt{v}-\varepsilon+\sqrt{-\varepsilon}+4\beta}{4(\eta-1)}+\frac{-2\sqrt{-\varepsilon}\sqrt{v}-4\beta\sqrt{-\varepsilon}+\varepsilon-\sqrt{-\varepsilon}-2\sqrt{v}-4\beta+v}{4\eta}\bigg).
\end{align}
\begin{table}[!htbp]
\centering
\begin{tabular}{||l|c|c|c||l|c|c|c||}\hline
$v$ & $\beta$ & $N$ &$\varepsilon_N$ & $v$ & $\beta$ & $N$ &$\varepsilon_N$ \\ \hline  
$29+8\sqrt{13}$ & $0$ &$0$& $~~-5.302~775~637~732_{(exact;0.178)}$ & $125+16\sqrt{61}$ & $0$ & $0$ &$-29.219~599~862~258_{(31;1521.157)}$ \\
$~$ & $1/2$ &$0$& $~-5.133~610~629~512_{(24;226.548)}$ & $~$ & $1/2$ & $0$ &$-29.215~374~513~860_{(exact;2.470)}$ \\
&$0$ &$1$ &$~-0.876~468~619~383_{( 27;555.165)}$& & $0$ &1&$-15.750~048~186~958_{(33; 2499.538)}$ \\
&$1/2$ &$1$ &$~-0.441~133~950~651_{( 27;522.314)}$& &$1/2$ &1&$-15.682~067~429~849_{(35;3706.651)}$ \\ 
&$~$ &$~$ &$~$& &$0$ &2&$~-6.329~594~513~877_{(36;15171.110)}$ \\
&$~$ &$~$ &$~$& &$1/2$ &2&$~-5.962~774~676~743_{(37;27567.614)}$\\
&$~$ &$~$ &$~$& &$0$ &3&$~-1.055~622~305_{(35;4565.332)}$ \\
&$~$ &$~$ &$~$& &$1/2$ &3&$~-0.560~462~417_{(35;4163.532)}$ \\ \hline
\end{tabular}\caption{The eigenenergies supported by the potentials $V(x)=-(29+8\sqrt{13})\,\sinh^4(z)/\cosh^6(z)$ and $V(x)=-(125+16\sqrt{61})\,\sinh^4(z)/\cosh^6(z)$, respectively. The subscript refer to the number of iterations and the computational times (in seconds) used by AIM.}\label{Tab4}
\end{table}

\noindent The eigenvalues evaluated using the roots of the termination condition \eqref{eq19} by means of the AIM sequences $\lambda_n(\eta)$ and $s_n(\eta), n=0,1,2,\dots $, initiated with $\lambda_0$ and $s_0$ as given by  equation \eqref{eq47} with $r_0$ is fixed at $r_0 =1/2$, are reported in Table \ref{Tab4}. In this table, $N$ refers to the state level not the number of possible nodes of the exact wave function.

\subsection{First-degree polynomial solutions}
\noindent In the case $N=1$, the first-degree polynomial solution reads, see Theorem \ref{Thm1}, 
\begin{align}\label{eq49}
f_1(\eta)&=1+\left(\dfrac{(4 \beta+2\sqrt{v}+\sqrt{-\varepsilon})(1+\sqrt{-\varepsilon})-v}{4 (1 + \sqrt{-\varepsilon})}\right)\eta,
\end{align}
subject to the following two constraints
\begin{align}\label{eq50}
7+4 \beta+2 \sqrt{-\varepsilon}-\sqrt{v}=0\Longrightarrow \varepsilon=-\dfrac{1}{4}(\sqrt{v}-7-4\beta)^2,\quad\quad for \quad v>(7+4\beta)^2,
\end{align}
and
\begin{align}\label{eq51}
v^2&-4 \left(2+\sqrt{-\varepsilon}\right)v^{3/2}-2 \left(\varepsilon-5 \sqrt{-\varepsilon}
+4 \beta \left(2+\sqrt{-\varepsilon}\right)-3\right)v\notag\\
&+4 \left(7+4 \beta \left(3+4 \sqrt{-\varepsilon}-\varepsilon\right)+11 \sqrt{-\varepsilon}+(-\varepsilon)^{3/2}-5 \varepsilon\right)\sqrt{v}\notag\\
&+16\,\beta^2 \left(3+4 \sqrt{-\varepsilon}-\varepsilon\right)+6\, \sqrt{-\varepsilon}+6\, (-\varepsilon)^{3/2}-11\varepsilon+\varepsilon^2-8\,\beta \left(\left(5+\sqrt{-\varepsilon}\right)\varepsilon-3-7 \sqrt{-\varepsilon}\right)=0.
\end{align}

\begin{figure}[!h]
\centering 
\includegraphics[width=3in, height=2.01in]{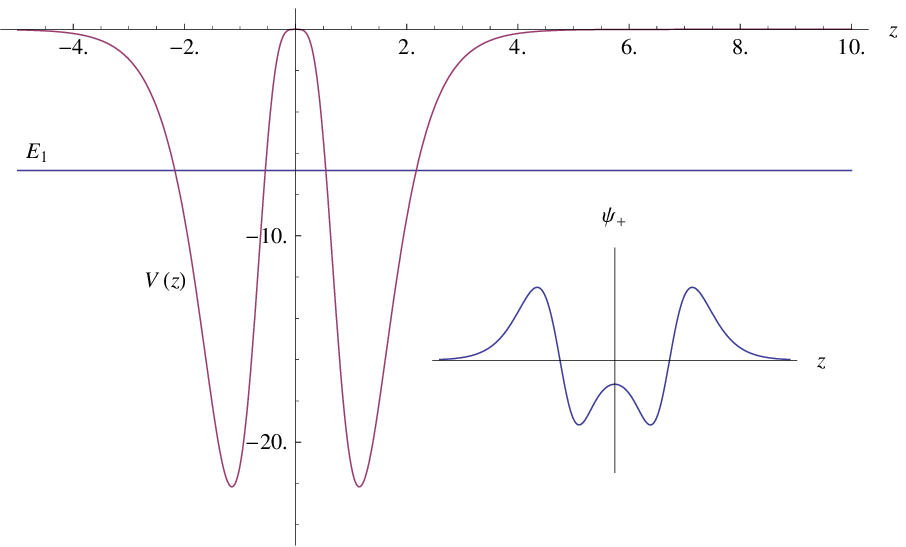}\qquad
\includegraphics[width=3in, height=2.0in]{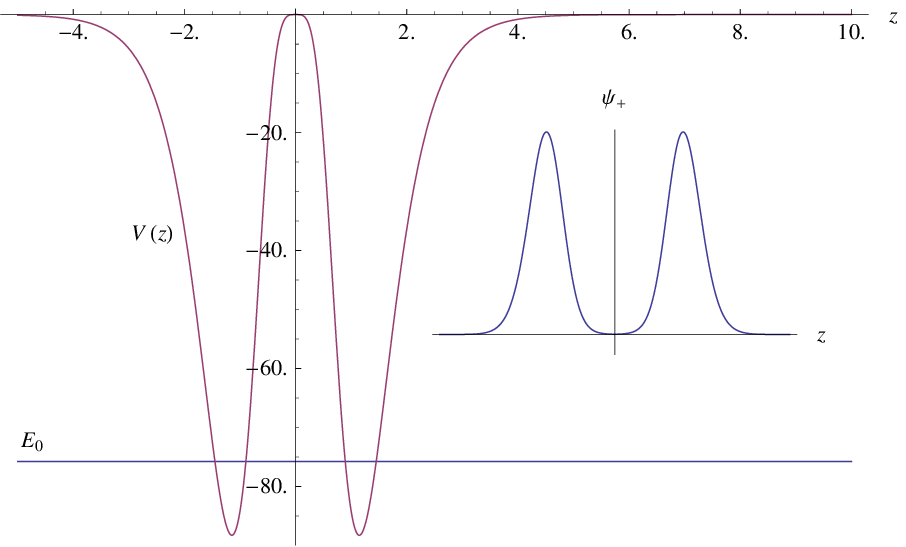}
\caption{Left: Plot of the potential $V(z)=-(149.574~256~933~312~\dots)\sinh^4(z)/\cosh^6(z)$  along with the exact eigenvalue and the exact wavefunction (Inset). Right: Plot of the potential $V(z)=-(595.838~654~872~035~\dots)\sinh^4(z)/\cosh^6(z)$  along with the exact eigenvalue and the exact wavefunction (Inset).}
\label{Fig3}
\vskip0.1true in
\includegraphics[width=3in, height=2.03in]{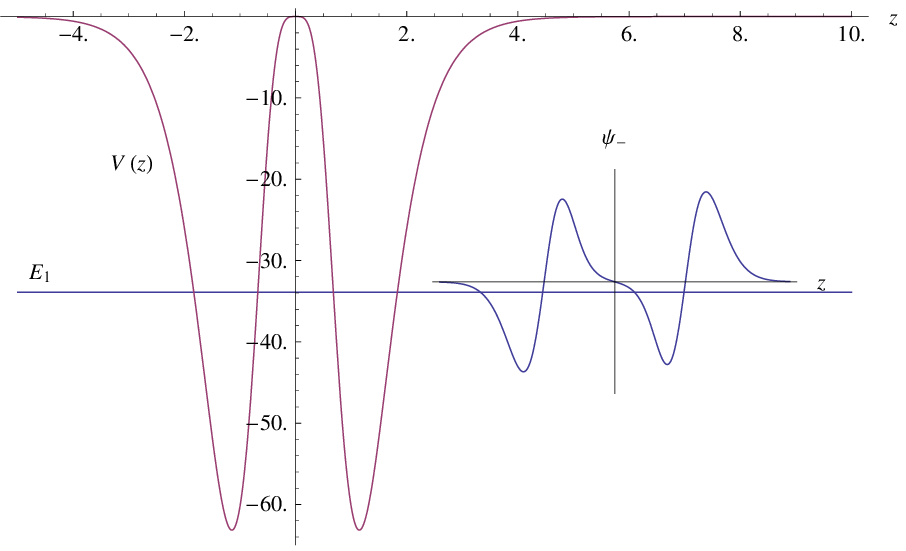}\qquad
\includegraphics[width=3in, height=2.03in]{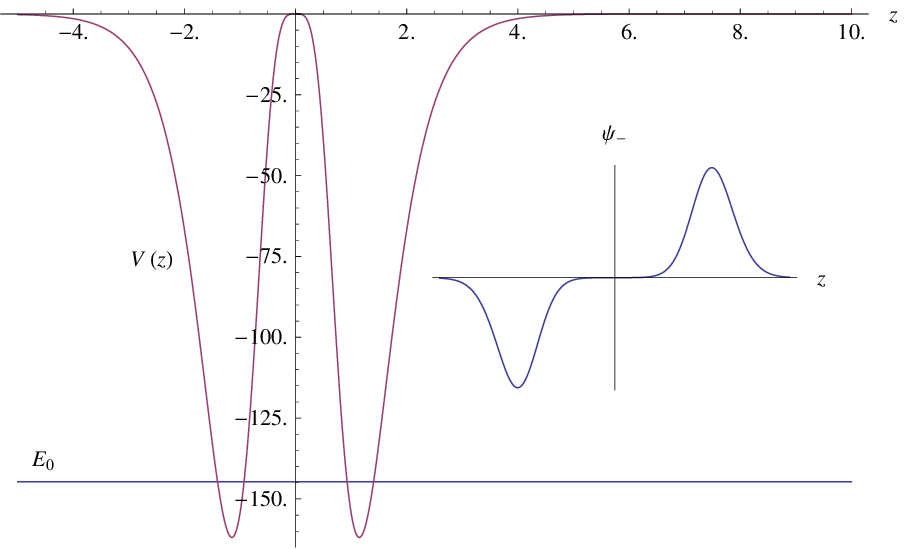}
\caption{Left: Plot of the potential $V(z)=-(426.232~048~\dots)\sinh^4(z)/\cosh^6(z)$  along with the exact eigenvalue and the exact wave function (Inset). Right: Plot of the potential $V(z)=-(1092.798~974~\dots)\sinh^4(z)/\cosh^6(z)$  along with the exact eigenvalue and the exact wave function (Inset)}
\label{Fig4}
\end{figure}

\noindent The non-zero solutions of this constraint system, for $\beta=0$, are 
\begin{align*}
v&= 149.574~256~933~312~63\dots, \qquad \varepsilon =-~ 6.838~370~069~149~139\dots,\\
v&= 595.838~654~872~035~2\dots,\qquad ~\varepsilon=- 75.775~340~860~142~68\dots,
\end{align*}
with exact wave functions
\begin{align}\label{eq52}
\psi_{+}(z)=e^{-6.115~027~737~739\dots \mbox{sech}^2(z)}\mbox{sech}^{2.615~027~737~739\dots}(z)(1 - 3.575~137~130~402\dots\times\mbox{sech}^2(z)),
\end{align}
and
\begin{align}\label{eq53}
\psi_{+}(z)=e^{-12.204~903~265~409~\mbox{sech}^2(z)} \mbox{sech}^{8.704~903~265~409\dots}(z)(1 -  0.967~778~541~967~\dots\times \mbox{sech}^2(z)),
\end{align}
respectively.  For $\beta=1/2$, the solution of the constraint system, \eqref{eq51} and \eqref{eq52}, gives
\begin{align*}
v&= ~426.232~048~026~951~5\dots, \qquad ~\varepsilon =-~33.903~765~749~927~9850\dots,\\
v&= 1092.798~974~117~571~6\dots,\qquad \varepsilon=-144.690~948~072~182~15\dots.
\end{align*}
with wave functions, respectively, given by
\begin{align}\label{eq54}
\psi_{-}(z)=\tanh(z)e^{-10.322~694~028~534\dots \text{sech}^2(z)} \text{sech}^{5.822~694~028~534~631\dots}(z) \left(1-3.339~804~928~329\dots \text{sech}^2(z)\right),
\end{align}
and
\begin{align}\label{eq55}
\psi_{-}(z)=
\tanh(z)e^{-16.528~755~050~803\dots \text{sech}^2(z)} \text{sech}^{12.028~755~050~803\dots}(z) \left(1-0.933~039~213~973\dots \text{sech}^2(z)\right).
\end{align}

\begin{table}[!htbp]
\begin{tabular}{|l|l|l||l|l|l|}
\hline
\multicolumn{3}{|c||}{$v= 149.574~256~933~312~63\dots$}&\multicolumn{3}{c|}{$v= 595.838~654~872~035~2\dots$} \\
\hline
$~$ & $n$ & $\varepsilon_n$& $~$ & $n$ & $\varepsilon_n$\\ \hline
\multirow{4}{*}{$\beta=0$} & $0$ & $-16.306~760~545~340_{(29;360.142)}$ & \multirow{4}{*}{$\beta=0$} & $0$ & $-75.775~340~860~141_{(Exact;0.371)}$ \\
 & $1$ & $-~6.838~370~069~149_{(Exact;0.371)}$ & & $1$ &$-52.875~071~648~495_{(36;896.369)}$\\
 & $2$ & $-~1.394~574~676~856_{(34;2000.300)}$&  & $2$ & $-33.784~057~730~949_{(38;4971.104)}$\\
 & $~$ & $~$&  & $3$ & $-18.715~749~915~481_{(42;11388.551)}$\\
  \hline
\multicolumn{3}{|c||}{$v= 426.232~048~026~951~5\dots$}&\multicolumn{3}{c|}{$v= 1092.798~974~117~571~6\dots$} \\
\hline
$~$ & $n$ & $\varepsilon_n$& $~$ & $n$ & $\varepsilon_n$\\ \hline
\multirow{4}{*}{$\beta=\frac12$} 
& $0$ & $-52.697~980~225~426_{(30;382.872)}$ & 
\multirow{4}{*}{$\beta=\frac12$} & $0$ & $-144.690~948~072~194_{(Exact;0.581)}$ \\
& $1$ & $-33.903~765~749~927_{(Exact;0.889)}$ &
 & $1$ & $-112.358~061~313~770_{(37;1118.803)}$ \\
 & $2$ & $-18.922~469~167~620_{(37;991.306)}$ & & $2$ & $-~83.768~650~791~018_{(41;2063.758)}$\\
 & $3$ & $-~7.902~902~698~027_{(41;1965.702)}$&  & $3$ & $-~59.031~914~212~835_{(45;4042.183)}$\\
 & $4$ & $-~1.273~518~290~111_{(44;3285.623)}$&  & $4$ & $-~38.275~969~150~953_{(48;6787.946)}$\\
  & $~$ & $~$&  & $5$ & $-~21.640~189~686~817_{(52;13098.384)}$\\  \hline
  \end{tabular}\caption{The discrete spectra supported by the potentials $V(z)=-149.574\dots\sinh^4(z)/\cosh^6(z), V(z)=-595.838\dots\sinh^4(z)/\cosh^6(z), V(x)=-426.232\dots\sinh^4(z)/\cosh^6(z),$ and $V(z)=-1092.798\dots\sinh^4(z)/\cosh^6(z)$. The subscripts refer to the number of iterations and the computational times (in seconds) used by AIM.}\label{Tab5}
  \end{table}

\noindent The plot of the exact wave functions \eqref{eq52}-\eqref{eq55} are displayed in Figures \ref{Fig3}-\ref{Fig4}. For each exact case, the rest of the discrete spectrum can be evaluated by using AIM and some of the eigenvalues are displayed in Table \ref{Tab5}. The accuracy of the eigenvalues to much higher number of decimal places can be obtain with some patience  specially for higher  iteration numbers for which the AIM computations may become tedious.

\subsection{Second-degree polynomial solution}

\noindent In the case $N=2$, the second-degree polynomial solutions for $\beta=0$ is 
\begin{align}
f_2(\eta)&=1+\dfrac{(2\sqrt{v}+\sqrt{-\varepsilon})(1+\sqrt{-\varepsilon})-v}{4 \left(1+\sqrt{-\varepsilon}\right)}\eta\notag\\
&+\dfrac{v^2-4(2+\sqrt{-\varepsilon})v^{3/2}+ 2 \left(5+7 \sqrt{-\varepsilon}-\varepsilon\right)v-4\sqrt{-\varepsilon}(\varepsilon-3\sqrt{-\varepsilon}-2)v^{1/2}-\sqrt{-\varepsilon}(6-11 \sqrt{-\varepsilon}+\varepsilon+\varepsilon\sqrt{-\varepsilon})}{32 (1+ \sqrt{-\varepsilon})(2+\sqrt{-\varepsilon})}\eta^2
\end{align}
subject to the exact values of $\varepsilon$ and $v$ given as: \\
\begin{align*}
(\varepsilon,v)&=(-235.972~747~187~322~915~\dots, 1740.792~785~280~901~099~\dots),\\
(\varepsilon,v)&=( -84.008~276~606~551~\dots, 860.319~634~232~780~644~\dots),\\ 
(\varepsilon,v)&=( -8.143~960~834~197~\dots, 279.141~396~634~685~463~\dots).
\end{align*}
\vskip0.2true in
\noindent For $\beta=1/2$, the second-order polynomial solution reads
\begin{align}
f_2(\eta)&=1+\dfrac{(2+2\sqrt{v}+\sqrt{-\varepsilon})(1+\sqrt{-\varepsilon})-v}{4 \left(1+\sqrt{-\varepsilon}\right)}\eta\notag\\
&+\dfrac{v^2-4(2+\sqrt{-\varepsilon})v^{3/2}+ 2 \left(1+5 \sqrt{-\varepsilon}-\varepsilon\right)v+4(4+\sqrt{-\varepsilon}(8-5\sqrt{-\varepsilon}-\varepsilon))v^{1/2}+24+10\sqrt{-\varepsilon}(5-\varepsilon)-\varepsilon(35-\varepsilon)}{32 (1+ \sqrt{-\varepsilon})(2+\sqrt{-\varepsilon})}\eta^2
\end{align}
subject to the exact values of $\varepsilon$ and $v$ as
\begin{align*}
(\varepsilon,v)&=( -349.620~385~570~634~540~\dots, 2539.784~747~349~247~690~\dots),\\
(\varepsilon,v)&=( -156.512~009~095~311~355~\dots, 1445.592~787~027~749~297~\dots),\\ 
(\varepsilon,v)&=(-8.143~960~834~197~590~\dots, 642.496~980~045~734~503~\dots ).
\end{align*}

\begin{figure}[!htbp]
\centering 
\includegraphics[width=2in, height=2.01in]{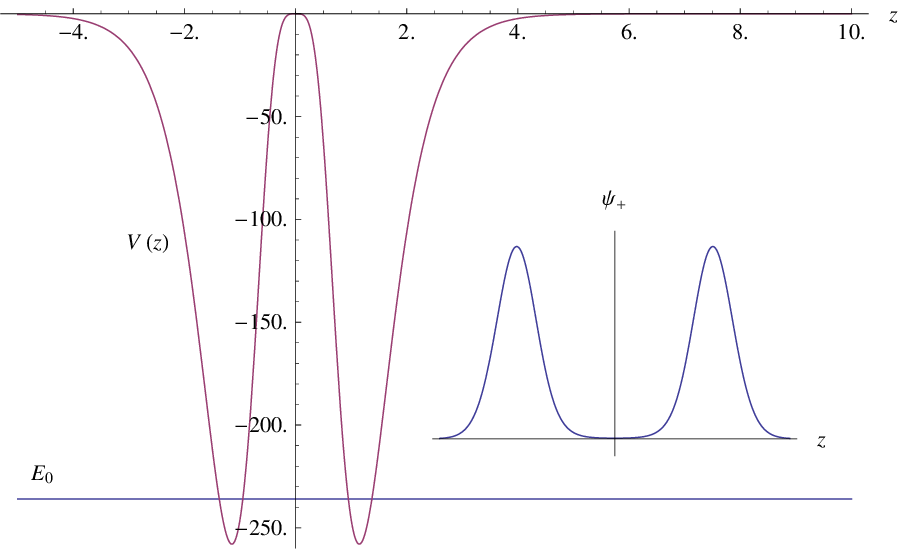}\quad
\includegraphics[width=2in, height=2.0in]{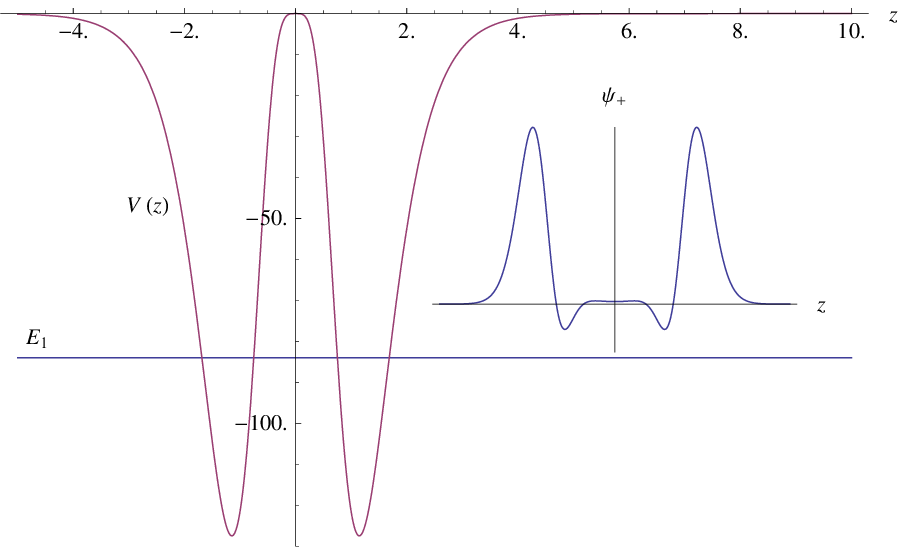}\quad
\includegraphics[width=2in, height=2.0in]{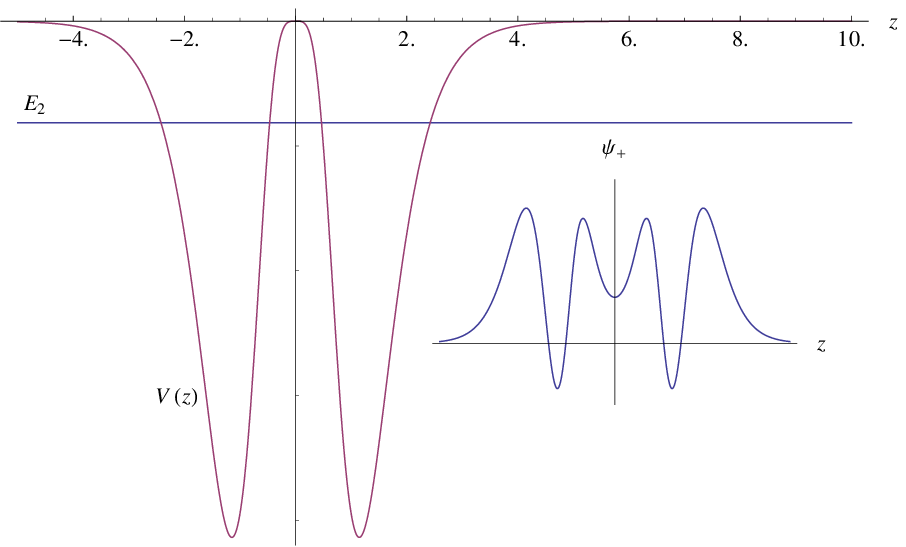}
\caption{Plot of the potentials $V(z)=-1740.792~\dots\sinh^4(z)/\cosh^6(z)$  $V(z)=-860.319~\dots\sinh^4(z)/\cosh^6(z)$  and $V(z)=-279.141~\dots\sinh^4(z)/\cosh^6(z)$, respectively,  along with the exact eigenvalue and the exact wave function (Inset).}
\label{Fig5}
\quad
\includegraphics[width=2in, height=2.01in]{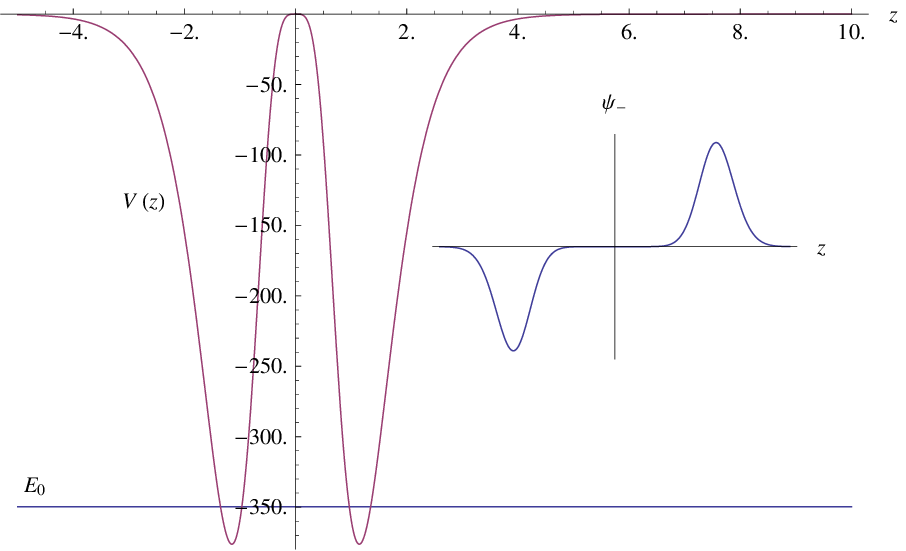}\quad
\includegraphics[width=2in, height=2.0in]{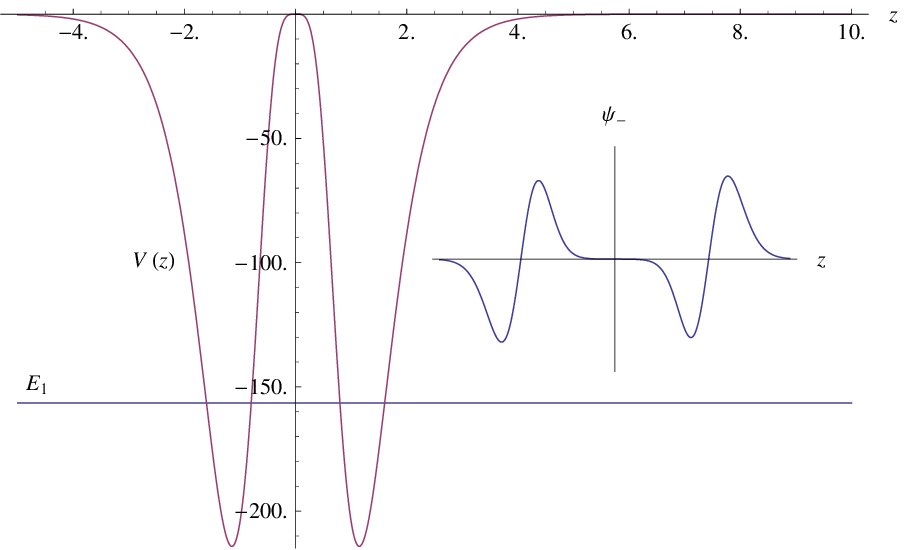}\quad
\includegraphics[width=2in, height=2.0in]{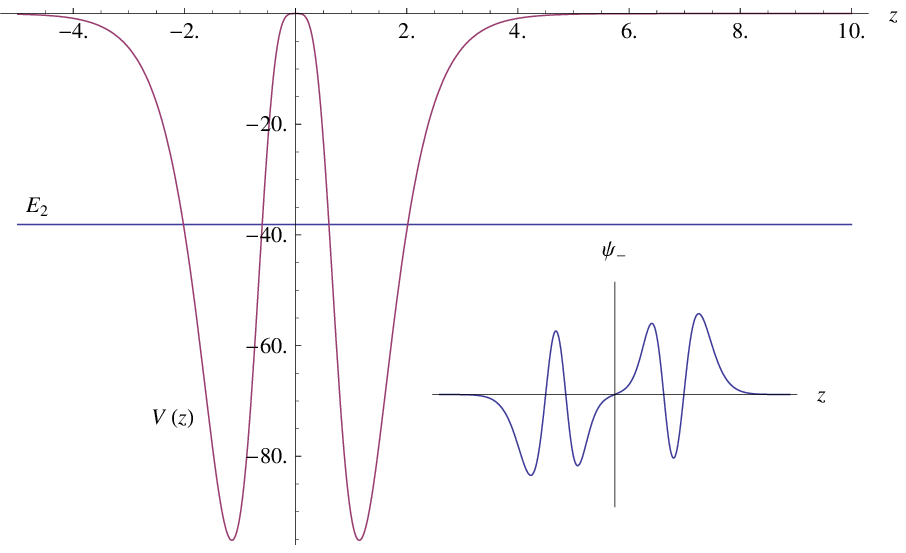}
\caption{Plot of the potentials $V(z)=-2539.784~\dots\sinh^4(z)/\cosh^6(z)$  $V(z)=-1445.592~\dots\sinh^4(z)/\cosh^6(z)$  and $V(z)=-642.496~\dots\sinh^4(z)/\cosh^6(z)$, respectively,  along with the exact eigenvalue and the exact wave function (Inset).}
\label{Fig6}\end{figure}

\noindent Plots of the exact wave functions are displayed in Fig. \ref{Fig5} and in Fig. \ref{Fig6} along with potential. For each exact case, the rest of the discrete spectrum can be approximated by the use of AIM, and some are displayed in Table \ref{Tab5}. The accuracy of the eigenvalues to a much higher number of decimal places  can be obtained by using more  iterations provided the numerical computing environment can support it.

\begin{table}[!htbp]
\begin{tabular}{|l|c|c||l|c|c||l|c|c|}
\hline
\multicolumn{3}{|c||}{$v= 1740.792~785~280~901~099~\dots$}&\multicolumn{3}{|c||}{$v= 860.319~634~232~780~644~\dots$} &\multicolumn{3}{|c|}{$v= 279.141~396~634~685~463~\dots$}\\
\hline
$~$ & $n$ & $\varepsilon_n$& $~$ & $n$ & $\varepsilon_n$& $~$ & $n$ & $\varepsilon_n$\\ \hline
\multirow{4}{*}{$\beta=0$} & $0$ & $~-235.972~747~187~323_{(Exact;1.522)}$ & \multirow{4}{*}{$\beta=0$} & $0$ & $-112.278~226~832~940_{(33;565.591)}$& \multirow{4}{*}{$\beta=0$} & $0$ & $-33.052~738~536~688_{(31;387.010)}$ \\
 & $1$ &$-194.196~436~729~193_{(39;1595.864)}$ & & $1$ & $~~-84.008~276~606~551_{(Exact;1.572)}$& & $1$ &$-18.591~111~216~762_{(34;535.603)}$ \\
 & $2$ & $-156.132~723~931~652_{(43;2893.056)}$&  & $2$ & $~-59.502~913~853~147_{(42;2323.202)}$& &$2$ & $~-8.143~960~834~198_{(Exact;1.709)}$\\
 & $3$ & $-121.862~017~411~137_{(46;4822.630)}$&  & $3$ & $~-38.901~417~419~466_{(46;4465.122)}$& & $3$&$~-1.902~068~729~808_{(40;1595.618)}$\\
  \hline
  \hline
\multicolumn{3}{|c||}{$v= 2539.784~747~349~247~690~\dots$}&\multicolumn{3}{|c||}{$v= 1445.592~787~027~749~297~\dots$} &\multicolumn{3}{|c|}{$v= 642.496~980~045~734~503~\dots$}\\
\hline
$~$ & $n$ & $\varepsilon_n$& $~$ & $n$ & $\varepsilon_n$& $~$ & $n$ & $\varepsilon_n$\\ \hline
\multirow{4}{*}{$\beta=\dfrac12$} & $0$ & $~-349.620~385~570~635_{(Exact;2.792)}$ & \multirow{4}{*}{$\beta=\dfrac12$} & $0$ & $-194.254~628~745~112_{(32; 399.421)}$& \multirow{4}{*}{$\beta=\dfrac12$} & $0$ & $-82.177~087~688~263_{(32;423.219)}$ \\
 & $1$ &$-298.395~312~298~514_{(38;1183.456)}$ & & $1$ & $-156.512~009~095~311_{(Exact;2.214)}$& & $1$ &$-58.252~330~895~293_{(36;800.408)}$ \\
 & $2$ & $-250.863~567~015~215_{(44; 3374.220)}$&  & $2$ & $-122.493~254~793~333_{(42;2423.644)}$& &$2$ & $-38.115~338~099~145_{(Exact;1.184)}$\\
 & $3$ & $-207.088~119~953~586_{(47;5641.614)}$&  & $3$ & $-92.289~274~418~582_{(46;4784.408)}$& & $3$&$-21.902~705~416~297_{(43;2738.455)}$\\
  \hline
  \end{tabular}\caption{The discrete spectra supported by the potentials $V(z)=-149.574\dots\sinh^4(z)/\cosh^6(z), V(z)=-595.838\dots\sinh^4(z)/\cosh^6(z), V(x)=-426.232\dots\sinh^4(z)/\cosh^6(z),$ and $V(z)=-1092.798\dots\sinh^4(z)/\cosh^6(z)$. The subscript refer to the number of iterations and the computational times (in seconds) used by AIM.}\label{Tab6}
  \end{table}

\section{Conclusion}\label{conc}

\noindent In this work, the exact and approximate solutions of Schr\"odinger's equation with various hyperbolic potentials \eqref{eq3}, $m=0,1,2$ are discussed.  For $m=2$, the corresponding 
Schr\"odinger equation admits polynomial solutions provided that certain constraints on the potential parameters are satisfied. A general existence theorem is devised that allows us to enumerate all such solutions of the confluent Heun equation: the proof of this theorem is presented in Appendix~I. This theory has the advantage  of easy implementation requiring only the computation of three-term recurrence relation. The theory is illustrated by the computation of the exact eigenvalues for the higher order polynomial solutions which are reported in Appendix~II. For non-polynomial cases AIM is also used to provide very accurate numerical approximations.

\section*{Acknowledgments}
\medskip
\noindent Partial financial support of this work under Grant Nos. GP3438 and GP249507 from the 
Natural Sciences and Engineering Research Council of Canada
 is gratefully acknowledged by us (respectively RLH and NS). 

\section*{Appendix I: The proof of Theorem \ref{Thm1}}
\begin{theorem}\label{Thm1}
The necessary condition for the existence of $N$-degree polynomial solutions of the differential equation
\begin{align*}
(a_2z^2+a_1z)f''(z)+(b_2z^2+b_1z+b_0)f'(z)-(\tau_1\, z+\tau_0)f(z)=0
\end{align*}
is \begin{align*}
\tau_1=N b_2,\qquad N=0,1,2,\dots.
\end{align*}
and the polynomial solutions are give explicitly by
\begin{align*}
f_N(z)=\sum_{k=0}^N \dfrac{P_k(\tau_0)}{k!\,a_1^k\left({b_0}/{a_1}\right)_k}\,z^k,
\end{align*}
where, for each $N$, the finite sequence of the polynomials $\{P_k(\tau_0)\}_{k=0}^N$ satisfies a three-term recurrence relation, for $0\leq k\leq n+1$,
\begin{align*}
P_{k+1}(\tau_0)=(\tau_0-k(k-1)a_2-kb_1)P_k(\tau_0)+kb_2(N-k+1)((k-1)a_1+b_0)P_{k-1}(\tau_0),
\end{align*}
initialized with $P_{-1}(\tau_0)=0,~P_0(\tau_0)=1$. 
 \end{theorem}
 
 \noindent{\bf Proof}\hfil\break\medskip

 \noindent  The differential equation
\begin{align*}
(a_2z^2+a_1z)f''(z)+(b_2z^2+b_1z+b_0)f'(z)-(\tau_1\, z+\tau_0)f(z)=0,
\end{align*}
with real constants $a_j,j=1,2$, $b_k,k=2,1,0$ has two regular singular points, namely at $z=0$ and $z=-a_1/a_2$, in addition to an irregular singular point at $z=\infty$. The domain of definition is $z\in (0,-a_1/a_2)$ if $a_2a_1<0$ or  $z\in (-a_1/a_2,0)$ if $a_2a_1>0.$ In the neighbourhood of the singular point $z=0$, the formal series solution takes the form $y(z)=\sum_{k=0}^\infty c_k z^k$ since the exponents of the regular point $z=0$ are $s=0$ and $s=1-b_0/a_1$. On substituting this expression for  $y(z)$ into the differential equation and employing all the necessary shifting of the summation indices, the recurrence relation for the coefficients $c_k$ reads
\begin{align*}
(k(k+1)a_1+(k+1)b_0)c_{k+1}+(k(k-1)a_2+kb_1-\tau_0)c_k+((k-1)b_2-\tau_1) c_{k-1}=0,
\end{align*}
for $k=0,1,2\dots,$ with the convention $c_{-1}=0,c_0=1.$ For an $N$th  degree polynomial solution, this linear system can be expressed as
 \begin{align*}
\sum_{j=0}^2 \left[(k-j)(k-j+1)a_{j+1}+(k-j+1)b_j-\tau_{j-1}\right]c_{k-j+1}=0
\end{align*}
where $a_{3}=\tau_{-1}=0$ and $k=0,1,\dots, N+1$. The system of the $N+2$ equations breaks down into three subclasses. For $k=N+1$, the necessary condition for polynomial solutions is 
 \begin{align*}
(Nb_2-\tau_1)c_N=0.
\end{align*}
The second subclass also consists of a single equation, namely $k=N$, that yields the sufficient condition as
\begin{align*}
(N(N-1)a_2+Nb_1-\tau_0)c_N+((N-1)b_2-\tau_1) c_{N-1}=0.
\end{align*}
Using $\tau_1=Nb_2$, it easily follows that
\begin{align*}
(N(N-1)a_2+Nb_1-\tau_0)c_N-b_2 c_{N-1}=0.
\end{align*}
The third subclass consists of $N$ equations given by $k=0,1,\dots,N-1$ and used to evaluate the polynomial coefficients $c_N$, for example, by Cramer's rule. Indeed, the $N$-equations can be written as
\begin{align*}
\begin{pmatrix}
\beta_1&0&0&0&0&\dots&0&0&0&0\\
\gamma_2&\beta_2&0&0&0&\dots&0&0&0&0\\
\eta_3&\gamma_3&\beta_3&0&0&\dots&0&0&0&0\\
0&\eta_4&\gamma_4&\beta_4&0&\dots &0&0&0&0\\
\vdots& \ddots&\ddots&\ddots&\ddots&\dots&\vdots&\vdots&\vdots&\vdots\\
0& 0&0&0&0&\dots&\eta_{N-1}&\gamma_{N-1}&\beta_{N-1}&0\\
0& 0&0&0&0&\dots&0&\eta_{N}&\gamma_{N}&\beta_N\\
\end{pmatrix} \begin{pmatrix} c_1\\ c_2\\ c_3\\ c_4\\ \vdots\\ c_{N-1}\\ c_N
\end{pmatrix}=\begin{pmatrix} \tau_0\\ \tau_1\\ 0\\ 0\\ \vdots\\ 0\\ 0
\end{pmatrix}\end{align*}
where
\begin{align*}
\beta_j&=j(j-1)a_1+jb_0,\qquad j=1,2,\dots,N,\\
\gamma_j&=(j-2)(j-1)a_1+(j-1)b_0-\tau_0,\qquad j=2,3,\dots,N-1,\\
\eta_j&=(j-2)b_2-\tau_1,\qquad j=3,4,\dots,N-2,
\end{align*}
The non-trivial solutions $c_k,k=1,2,\dots,N$ require that $\prod_{j=1}^N
\beta_j\neq 0$  which yields $N!a_1^N(b_0/a_1)_N\neq 0$. The application of Cramer's rule allows us to to express the polynomial solutions of the differential equation as
\begin{align*}
f_N(z)=\sum_{k=0}^N \dfrac{P_k(\tau_0)}{k!\,a_1^k\left({b_0}/{a_1}\right)_k}\,z^k,
\end{align*}
where, for each $N$, the finite sequence of the polynomials $\{P_k(\tau_0)\}_{k=0}^N$ satisfies a three-term recurrence relation, for $0\leq k\leq n+1$,
\begin{align*}
P_{k+1}(\tau_0)=(\tau_0-k(k-1)a_2-kb_1)P_k(\tau_0)+kb_2(N-k+1)((k-1)a_1+b_0)P_{k-1}(\tau_0),
\end{align*}
utilized with $P_{-1}(\tau_0)=0,~P_0(\tau_0)=1$. The first few  degree-$N$ polynomial solutions of the differential equation are
\begin{align*}
(a_2z^2+a_1z)f''(z)+(b_2z^2+b_1z+b_0)f'(z)-(\tau_1\, z+\tau_0)f(z)=0
\end{align*}
\begin{itemize}
\item $N=0$: the constant solution is 
\begin{align*}f_0(\eta)=1,
\end{align*} subject to $\tau_1=0$ and $\tau_0=0$.

\item $N=1$: the first degree solution is 
\begin{align*}f_1(\eta)=1+\dfrac{\tau_0}{b_0}\eta,
\end{align*} subject to $\eta_1=b_2$ and $\tau_0^2-b_1\tau_0+b_0b_2=0$.

\item $N=2$: the second degree solution is 
\begin{align*}
f_2(\eta)&=1+\dfrac{\tau_0}{b_0}\eta+\dfrac{\tau_0^2-b_1\tau_0+2b_0b_2}{2b_0(a_1+b_0)}\eta^2
\end{align*} 
subject to 
\begin{align*}
\tau_1&=2b_2,\\
\tau_0^3&- (2 a_2+3 b_1)\tau_0^2+2 (b_1 (a_2+b_1)+(a_1+2 b_0) b_2)\tau_0-4b_0 (a_2+b_1)b_2=0.
\end{align*}

\item $N=3$: the third-degree solution is
\begin{align*}
f_3(\eta)&=1+\dfrac{\tau_0}{b_0}\eta+\dfrac{\tau_0^2-b_1\tau_0+3b_0b_2}{2b_0(a_1+b_0)}\eta^2+
\dfrac{\tau_0^3-(2a_2+3b_1)\tau_0^2-(2a_2b_1 +2 b_1^2+4a_1b_2+7 b_0b_2) \tau_0-6 (a_2+b_1)b_0 b_2}{6b_0 (a_1+b_0) (2a_1+b_0)}\eta^3,
\end{align*}
subject to
\begin{align*}
\tau_1&=3b_2,\\
\tau_0^4&-2 (4 a_2 + 3 b_1) \tau_0^3+(12 a_2^2 + 26 a_2 b_1 + 11 b_1^2 + 10 (a_1 + b_0) b_2) 
\tau_0^2\\
&-6 (b_1 (a_2+b_1) (2 a_2+b_1)+(4 a_1 a_2+8 a_2 b_0+3 a_1 b_1+5 b_0 b_1) b_2)\tau_0+9 b_0 b_2 (2 (a_2+b_1) (2a_2+b_1)+(2a_1+b_0) b_2)=0.
 \end{align*}

\item $N=4$: the fourth-degree solution is
\begin{align*}
&f_3(\eta)=1+\dfrac{\tau_0}{b_0}\eta+\dfrac{\tau_0^2-b_1\tau_0+4b_0b_2}{2b_0(a_1+b_0)}\eta^2+
\dfrac{\tau_0^3-(2a_2+3b_1)\tau_0^2-(2a_2b_1 +2 b_1^2+6a_1b_2+10 b_0b_2) \tau_0-8 (a_2+b_1)b_0 b_2}{6b_0 (a_1+b_0) (2a_1+b_0)}\eta^3\notag\\
&+\dfrac{\xi}{24 b_0 (a_1 + b_0) (2 a_1 + b_0) (3 a_1 + b_0)}\eta^4,
\end{align*}
where
\begin{align*}
\tau_0^4&-2(4 a_2 + 3 b_1) \tau_0^3+(12 a_2^2 + 26 a_2 b_1 + 11 b_1^2 + 18 a_1 b_2 + 16 b_0 b_2) \tau_0^2\\
&-2 (6 a_2^2 b_1 + 9 a_2 b_1^2 + 3 b_1^3 + 18 a_1 a_2 b_2 + 34 a_2 b_0 b_2 + 
   15 a_1 b_1 b_2 + 22 b_0 b_1 b_2)\tau_0\\
   &+24 b_0 b_2 (2 a_2^2 + 3 a_2 b_1 + b_1^2 + 2 a_1 b_2 + b_0 b_2)=0
   \end{align*}
subject to
\begin{align*}
\tau_1&=4b_2,\\
\tau_0^5&-10 (2 a_2 + b_1) \tau_0^4+(108 a_2^2 + 130 a_2 b_1 + 35 b_1^2 + 30 a_1 b_2 + 20 b_0 b_2) 
\tau_0^3\\
&-2 (72 a_2^3 + 186 a_2^2 b_1 + 127 a_2 b_1^2 + 25 b_1^3 + 138 a_1 a_2 b_2 + 
   134 a_2 b_0 b_2 + 69 a_1 b_1 b_2 + 60 b_0 b_1 b_2)\tau_0^2\notag\\
   &+8 (18 a_2^3 b_1 + 33 a_2^2 b_1^2 + 18 a_2 b_1^3 + 3 b_1^4 + 54 a_1 a_2^2 b_2 + 
   108 a_2^2 b_0 b_2 + 66 a_1 a_2 b_1 b_2 + 110 a_2 b_0 b_1 b_2 \notag\\
   &+ 
   18 a_1 b_1^2 b_2 + 26 b_0 b_1^2 b_2 + 9 a_1^2 b_2^2 + 24 a_1 b_0 b_2^2 + 
   8 b_0^2 b_2^2)\tau_0-32 b_0 b_2 (18 a_2^3 + 33 a_2^2 b_1 + 18 a_2 b_1^2 + 3 b_1^3\\
   & + 21 a_1 a_2 b_2 + 
   10 a_2 b_0 b_2 + 9 a_1 b_1 b_2 + 4 b_0 b_1 b_2)=0.
 \end{align*}
 \end{itemize}  

\section*{Appendix II: High-order polynomial solutions for $m=2$}
\noindent For the third-degree polynomial solution $N=3$, in the symmetric case $\beta=0$, we have the exact eigenvalue corresponding the potential strength as follows: 
\begin{align*}
(\varepsilon, v)&=( -485.633~694~118~171~701~\dots, 3489.760~664~445~471~178~\dots),\\
(\varepsilon,v)&=
(-251.403~743~430~891~744~\dots, 2181.957~959~376~591~216~\dots),\\
(\varepsilon,v)&=
(-91.612~310~090~595~802~\dots, 1165.735~158~983~190~844~\dots),\\
(\varepsilon,v)&=
(-9.314~641~090~299~513~\dots,  445.377~945~981~708~794~\dots),
\end{align*}
while for the anti-symmetric case, $\beta=1/2$, the exact pair of the eigenvalues and the potential strength are
\begin{align*}
(\varepsilon,v)&=(-644.012~578~551~987~881~\dots, 4590.713~712~215~915~292),\\
(\varepsilon,v)&=
(-368.672~229~400~112~178~\dots, 3069.345~989~249~095~172~\dots),\\
(\varepsilon,v)&=(
-167.611~705~221~567~520~\dots,  1839.808~408~499~296~398~\dots),\\
(\varepsilon,v)&=(-41.996~677~058~480~376~\dots,897.659~642~186~544~995~\dots).
 \end{align*}
\noindent For the fourth-degree polynomial solution $N=4$, in the symmetric case $\beta=0$, we have the exact eigenvalue corresponding the potential strength as follows: 
\begin{align*}
(\varepsilon,v)&=(-824.756~984~779~279~158~\dots, 5842.640~214~370~125~564~\dots),\\
(\varepsilon,v)&=(-508.312~652~989~564~072~\dots, 4107.730~662~518~437~769~\dots),\\ 
(\varepsilon,v)&=(-266.051~362~321~527~012~\dots, 
2664.847~602~800~234~089~\dots),\\
(\varepsilon,v)&=(
-98.740~362~436~827~665~\dots, 1511.159~657~792~798~942~\dots),\\
(\varepsilon,v)&=(
-10.392~841~434~278~337~\dots, 647.579~635~239~168~471~
\dots).
\end{align*}
while for the anti-symmetric case, $\beta=1/2$, the exact pair of the eigenvalues and the potential strength are
\begin{align*}
(\varepsilon,v)&=(-1027.866~880~025~139~082~\dots,  7245.538~017~777~396~086~\dots),\\
(\varepsilon,v)&=(-670.322~583~106~494~693~\dots, 
5297.099~783~021~461~760~\dots),\\
(\varepsilon,v)&=(
-386.896~254~447~370~708~\dots, 3640.838~016~388~864~321~\dots),\\
(\varepsilon,v)&=
(-178.134~857~500~181~954~\dots, 2274.663~666~881~406~389~\dots),\\
(\varepsilon,v)&=(-45.630~488~564~892~802~\dots, 1190.944~836~521~014~392~\dots).
\end{align*}
\noindent For the fifth-degree polynomial solution $N=5$, in the symmetric case $\beta=0$, we have the exact eigenvalue corresponding the potential strength as follows: 
\begin{align*}
(\varepsilon,v)&=(-1253.342~243~407~552~930~\dots, 8799.405~778~488~506~227~\dots),\\
(\varepsilon,v)&=(-854.700~662~342~345~719~\dots, 6637.446~939~245~948~353~\dots),\\
(\varepsilon,v)&=
(-530.130~397~260~421~230~\dots, 4767.781~177~102~649~014~\dots),\\
(\varepsilon,v)&=(
-280.049~454~485~697~568~\dots, 3188.788~212~420~147~429~\dots),\\
(\varepsilon,v)&=(
-105.490~706~285~346~518~\dots, 1895.882~577~289~245~947~\dots),\\
(\varepsilon,v)&=(
-11.402~086~879~730~237~\dots, 
885.264~529~414~657~142~\dots),
 \end{align*}
while for the anti-symmetric case, $\beta=1/2$, the exact pair of the eigenvalues and the potential strength are
\begin{align*}
(\varepsilon,v)&=(-1501.183~061~037~531~656~\dots,10504.242~614~555~052~587~\dots),\\
(\varepsilon,v)&=(
-1061.446~073~598~550~701~\dots, 8128.768~446~594~073~010~\dots),\\
(\varepsilon,v)&=(
-695.745~365~800~573~168~\dots, 
6045.680~016~488~041~376~\dots),\\
(\varepsilon,v)&=(-404.415~553~322~089~445~\dots,4253.670~799~351~802~332~\dots),\\
(\varepsilon,v)&=(-188.182~487~408~993~217~\dots,2749.526~171~415~990~809~\dots),\\
(\varepsilon,v)&=(
-49.069~088~680~629~925~\dots, 
1521.769~671~468~867~580~\dots).
\end{align*}

\end{document}